# A study of strategy to the remove and ease TBT for increasing export in GCC6 countries

Yong-Jae Kim[1]

[1]Department of Business Administration, Korea Polytechnic University, Prof. ph.D.

yjkim@kpu.ac.kr

■ *Abstract*

The last technical barriers to trade(TBT) between countries are Non-Tariff Barriers(NTBs), meaning all trade barriers are possible other than Tariff Barriers. And the most typical examples are (TBT), which refer to measure Technical Regulation, Standards, Procedure for Conformity Assessment, Test & Certification etc. Therefore, in order to eliminate TBT, WTO has made all membership countries automatically enter into an agreement on TBT. The purpose of this paper is to analyze and investigate to find out the appropriate ways to remove/reduce TBT to export the goods between Korea and GCC 6countries. The elimination strategy of TBT with aid of technical regulations or standards is beyond this paper scope and only the conformity assessment shall be considered as the strategic measure of eliminating TBT. This paper of scope to accord with the international standards corresponding to countries technical regulations and standards, every membership countries most present to prevent TBT related Specific Trade Concern (STC) to WTO. This paper analyzes International rule & system with following research methodology. First, the paper made use of reviewing 2nd data analysis and focused on group interview, and then compared with analysis on the international rule and system such as customs system, technical regulation, and standard through test &certification procedure and inspection. Second, this paper reviewed electric/electronic test, certification, and calibration. Third, through analyzing MRA between Korea and GCC 6countries the paper focused on SDOC, Mutual Acceptance of International test & certification and IEC CB scheme. The result shows that it is important to promote multi-track agreements with the countries that need a short-term promotion. Moreover, it is necessary to conclude an agreement with the leading countries in the Middle East and GCC 6countries. This paper intends to draw a conclusion and make implication as follows. Through the reviews and analysis, we can see the importance of promotion in FTA and MRA with the efforts to remove/reduce the obstacles and promote to make a remove/reduce TBT by MRA between Korea and GCC 6countries.

■ *Keyword*

FTA, Standards, Conformity Assessment, TBT(Technical barrier to trade), MRA, WTO

# 1. INTRODUCTION

This paper purpose make to remove and to ease TBT of industrial products such as IT, S/W, IOT, BigData, Home network etc. Research methodology is review 2nd data analysis and focus group Interview Government officer, Professor and CEO.

This paper compare & analyze International rule & system as follow.

First, It is to compare & analyze the standard, technical regulation, Test & certification procedure and Inspection. Second, it is review electric/electronic Test, certification and calibration. Third, it is analyze MRA between Korea and GCC 6countries(GCC Member: Saudi Arabia, Kuwait, United Arab Emirates, Qatar, Oman, Bahrain.), SDoC, Mutual Acceptance of International certification such as ILAC, APLAC and IEC CB scheme.

Methods for the study are as follows. First, This paper analyzes the earlier studies designed to eliminate/remove/reduce TBTs. Second, This paper has used primary materials collected through the questionnaire-based survey to identify the difficulties and problems involved in the Korean companies exporting to GCC 6countries. Third, This paper analyzes previous research papers and other data on eliminating or reducing TBTs. Fourth, as far as the creation and implementation of government policies are concerned, This paper has used the data on the projects that provide financial supports for the Korean exporters with regard to the costs for acquiring international accreditation standards through the organizations affiliated with the Korean government.

Suggestions of the current study include ways to eliminate and reduce TBTs. First, one can resort to the certification with ILAC(International Laboratory Accreditation Cooperation), APLAC(Asia Pacific Laboratory Accreditation Cooperation) and the IEC System of Conformity Assessment Schemes for Electro technical Equipment and Components (IECEE) CB Scheme. Second, Korea operates an international accreditation in IT among twenty-three APEC member countries, which is also used with GCC 6countries. Third, we should help the developing countries in GCC 6countries commands the strategy to go around TBTs by supporting their creation and operation of testing and accrediting agencies through official development assistance.

This paper intends to draw conclusion and make implication as follows.

First, we must promote FTA and MRA. Second, we make to remove and to easy TBT by MRA between Korea and GCC 6countries. Although the MRA is a system where all parties that have concluded agreement enjoy the advantage, Korea is under a state of concluding only the stage 1 agreements(test documents) with GCC 6 countries. Also, we must conclude MRA stage 2agreements (certification documents)with GCC 6 countries.

# 2. PREVIOUS STUDIES

## 2.1 Strategy to remove & ease TBT in OECD

SDoC has strengths in cost reduction, time saving and product information protection aspects compared to the certification system while having vulnerability in terms of product safety issue, etc. Therefore, an effective post market surveillance of the regulation authorities must be supported to be operated effectively. WTO's TBT Committee has suggested that the SDoC is more effective TBT elimination method than the MRA (OECD, 2000).

## 2.2 Strategy to remove & ease TBT in APECTEL MRA

Testing and certification are expensive procedures for exporters, importers and regulators that increase the cost to users and delays the availability of products in a large number of markets.

All stakeholders benefit from simplified procedures that can reduce these costs. At the same time, regulators need to have confidence in the quality of testing that provides the basis for certification of equipment.

In June 1998, the APEC1 Telecommunications and Information Ministers agreed to streamline APEC-wide processes for the testing and type-approval of telecommunications equipment.

This landmark arrangement, the Mutual Recognition Arrangement for Conformity Assessment of Telecommunications Equipment (APEC TEL MRA2), was the first multilateral agreement of its type in the world.

This Arrangement streamlines the Conformity Assessment Procedures for a wide range of telecommunications and telecommunications-related equipment and facilitates trade among the APEC member economies.

It reduces a significant barrier to what is projected to be a US$60 billion industry by 2010. Its scope includes all equipment subject to telecommunication regulations, including wireline and wireless, terrestrial and satellite equipment. For such equipment, the MRA covers electromagnetic compatibility (EMC), specific absorption rate (SAR) and electrical safety aspects as well as purely telecommunications aspects of the conformity assessment requirements.

## 2.3 Previous Study

The earlier study that analyzed the effects that TBTs exerted on the Indian exports led to the conclusion of thirteen Regional Trade Agreements (RTA) with twenty-five countries in late 2014, thus contributing to boosting exports through lowering and abolishing tariffs (Lee Woong et al., 2014). The trade effects of TBTs are trade creation and trade restrictions, and creating and implementing policies for trade creation can contribute to increasing exports (Jang Youn-jun et al., 2011).

One way to increase exports by avoiding the non-tariff TBTs is as follows. If two countries sign MRA, Korea can export to the other without being certified by an accrediting agency of the other country and instead by getting certified by an accrediting agency in Korea. Thus, shortened period and reduced cost of certification can increase exports (Kim Yong-Jae, 2016)

## 3. THE SURVEY AND ANALYSIS OF TBTS NOTIFIED BY WTO

### 3.1 The Survey and Analysis of TBTs Notified by WTO

The number of the WTO's STCs related to TBTs was just 4 in 1995 when WTO launched, while steadily increased to 94 in 2012, 85 in 2014, and 86 in 2015. Between 1995 and 2015 total 885 specific trade concerns (STCs) were discussed. Such an increase in STCs constitutes a new trade restriction and thus a protectionist measure. The number of regular notifications that WTO members sent to Office for the WTO Agreement on Technical Barriers to Trade (WTO/TBT) in 2014 and 2015 were each 1,564 and 1,466 down from 1,630 in 2013.

### 3.2 TBT Statistics by Country

Of the ranking countries that issued most notifications between 1995 and 2015, the U

S was top (with 2,459 cases), Brazil was 2nd (with 1,331 cases), the EU was 3rd (with 1,253 cases), and Ecuador 8th (with 839 cases), Mexico 9th (with 813 cases), and Korea 11th. In 2015, of the ranking countries that issued most notifications, the US was top (with 283 cases), Ecuador was 2nd (with 126 cases), and Brazil was 3rd (with 119cases), while Korea issued80 notifications.

### 3.3 The survey and analysis of strategy to remove/ ease of the TBT

CEOs of small & medium companies located at Sihwa and Banwol industrial complex in Korea, which are exporting to 6 GCC countries in the Middle East, expect that their exports will increase by at least 20~30% more than now if FTA and MRA are concluded with 6 GCC countries. In addition, they want the government to actively support to obtain Halal Certification in order to explore the Middle East Islam market.1

As problems which should be addressed urgently among difficulties and problems they are facing in running small & medium companies exporting to 6 GCC countries in Sihwa and Banwol or trading, they pointed out difficulties with funding operation (50%) followed by marketing strategy (43%), weak exports (23.3%) and weak domestic demand/excessive inventory (13%) in order.

**[Table 1] Difficulties and problems in management(trade) in exporting to the Middle East(multiple responses)**

| Item / Response | Weak exports | Weak domestic demand/excessive inventory | Difficulties with funding operation | Difficulties with securing raw materials | Heavy taxation | Marketing strategy | Others | No response |
|---|---|---|---|---|---|---|---|---|
| No. of respondents | 7 | 4 | 15 | 0 | 1 | 13 | 1 | 2 |
| % of respondents | 23.3% | 13.3% | 50.5% | 0.0% | 3.3% | 43.3% | 3.3% | 6.7% |

## 4. STRATEGY TO REMOVE/ EASE OF THE TBT

TBT is an abbreviation for Technical Barriers to Trade' while this stands for the various obstacles in terms of trade that hinder the free movement of goods and services as the trading partner countries adopt and apply different Technical Regulations, Standards, Test &Certification Procedures and Inspection Systems, etc.

### 4.1 Strategy to Remove/Ease of the TBT

---

[1]This survey was carried out on 50 CEOs of small & medium companies located in Sihwa and Banwol in Korea from Feb.1 to March 5, 2018 and was analyzed by 50 experts in industry and academics through 3 rounds of discussions (F.G. I Focus Group Interview).

TBT makes the countries to harmonize technical regulations, standards or conformity assessments with the international standards and does not occur in case of being transparent. However, the fact is that TBT occurs if a specific country does not comply with the principles above during legislation and amendment of the laws related to technical regulations, standards or conformity assessments while STC must be submitted to settle this TBT. In the conformity assessment of ICT section, various methods of solution exist on TBT depending on the issue other than filing a lawsuit to WTO if a specific country operates the conformity assessment section differently from TBT.

**(1) Request for Introduction of SDoC System**

SDoC system stands for the one to guarantee market autonomy and raise efficiency of restriction as a system for the supplier to guarantee by evaluating whether its own product is appropriate for the concerned standard by escaping from the compulsory certification system which requires certification in relation to the product manufacture. Since SDoC(Supplier's Declaration of Conformity) is a follow-up and legal system, it is the method of releasing new products under the manufacturer's own responsibility to become responsible for various problems to follow.

Therefore, this system is the one that must be prepared with insurance system and product liability, etc. as well as social trust.

for example, Motor car(Auto car Industry), Handphone(Cellular phone)

**(2) Strategy of MRA**

The manufacturers of industrial products are able to export only after acquiring a compulsory standard certification mark. While MRA is concluded in order to save cost and time required for this, only the test report implemented at the exporting country is recognized if MRA stage 1(Test documents) is concluded while both the test report and the certification market may be implemented at the exporting country may be implemented if MRA stage 2(certification documents) is concluded.

If both countries conclude the MRA such as FTA, it would be opening the homeland market to the manufacturer of partner country since it is customs-free.

**4.2 Domestic Electric & Electronic Certification System and Related Laws**

- Conformity Assessment System of Korea

Supplier's Declaration of Conformity (SDoC) the one to guarantee market autonomy and raise efficiency of restriction as a system for the supplier to guarantee by evaluating whether its own product is appropriate for the concerned standard by escaping from the conventional compulsory certification system which requires certification in relation to the product manufacture.

**(1) Acceptance of Internationally Certified Test Report**

In addition to the method of concluding an MRA, various methods to recognized the test reports estimated at the partner country or a third country exist. Among them, the most widely used method is the one to accept test reports of the testing agencies that have been recognized by ILAC, APLAC and IEC CB Scheme. Test & Certification Based Infrastructure Setup Support.

**4.3 Comparative Analysis and strategy of remove and ease TBT**

The systems mentioned above have different characteristics from each other. If the comparative analysis is performed from the perspectives of scope of effect, intensity of effect and usage status in Korea, they can be summarized as follows.

**[Table 2] Comparative Analysis of Characteristics between TBT distribution Tools**

|  | Mutual recognition agreement (MRA) | (Supplier's Declaration of Conformity (SDoC) | MRA among AEOs | Project for overseas accreditation | Mutual acceptance of international certificates | | Project for creating accrediting infrastructure |
|---|---|---|---|---|---|---|---|
|  |  |  |  |  | ILAC (APLAC) | CB Scheme |  |
| Scope of effect | Agreement partner | All countries | Agreement partner | Agreement partner | Participating countries | Participating countries | Beneficiary countries (developing countries) |
| Intensity of effect | Progressive 100% | Limited to applicable products | Progressive 100% | Progressive 100% | By applicable area | By applicable area | Varying with levels of support |
| Current application to Korea | Phase 1 signed with 5 countries, negotiating with several countries | Applied to low-hazard products | Progressive 100% Facilitated customs clearance in importing countries | Facilitated customs clearance in importing countries | Civilian standards | Not used in EMC | Applying |
| Remarks | Phase 2 needs to be implemented with more countries | Application to more products in the short term Small need | Facilitated customs clearance in importing countries | Facilitated customs clearance in importing countries | Flexible response to varying foreign acceptance | Flexible response to varying foreign acceptance | Increased support needed |

Table compiled from various sources

If both countries conclude the MRA such as FTA, it would be opening the homeland market to the manufacturer of partner country since it is customs-free.

Among APEC members, time and cost required for preparing the copy of agreement can be saved if the MRA Guide prepared by this organization is used. Although Korea has concluded MRA stage 1 with the United States, Canada, Chile and Vietnam, etc., the effect is clearly shown only in the MRA with the United States.

## 4. CONCLUSION

This study intends to draw conclusions and make policy implications as follows.

First, we must promote a multi-track simultaneous agreements with the countries that have necessity of short-term promotion.

Second, the countries with necessity of short-term promotion on the preferential basis are China, Japan and USA, etc.

Third, it is necessary to conclude MRA agreement with the leading countries among the GCC 6 countries on the preferential basis. It is necessary to prepare negotiation on the preferential basis with GCC 6 countries

Fourth, support on the countries that have not fully prepared the conformity assessment system needs to be gradually extended. However, the method of support on these countries also must vary depending on the country. KTC(Korea Testing Certification) written MRA GCC test& certification company in 2015 years. So Korea exporting company can export to GCC goods with attached test & certification documents in Korea. To conclude, it is suggest MRA for the remove and reduce TBT to increase export and import among countries.

In its final analysis, I suggest MRA for the remove and reduce TBT to increase export and import among countries.

**Authors**

ISO TC 68 Member

ISO TC 195 Member

ISO/IEC SC32 WG1 Vice convenor

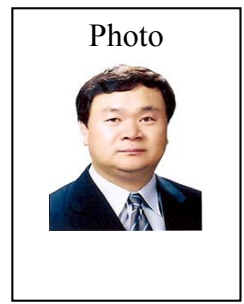